\documentclass[prd,showpacs,nofootinbib]{revtex4}

\usepackage{graphicx}

\usepackage{bm}
\newcommand{\vc}[1]{\bm{#1}}

\newcommand{\vf}[1]{\overrightarrow{#1}}

\newcommand{\ieps}[1]{\includegraphics[height=3in]{#1}}

\begin{document}

\title{Power filters for gravitational wave bursts: network operation for source position estimation}
\author{Julien Sylvestre}
\affiliation{LIGO Laboratory, California Institute of Technology,\\MS 18-34, Pasadena, CA 91125, USA.}

\email{jsylvest@ligo.caltech.edu}

\date{\today}

\begin{abstract}
A method is presented to generalize the power detectors for short bursts of gravitational waves that have been developed for single interferometers so that they can optimally process data from a network of interferometers.
The performances of this method for the estimation of the position of the source are studied using numerical simulations.
\end{abstract}


\maketitle

\section{introduction}
The most interesting advances in gravitational wave (GW) astronomy are likely to come from the cooperative operation of the interferometric detectors of gravitational radiation that are currently in an advanced stage of commissioning in North America, Europe, and Asia.
The two principal advantages of operating the interferometers in a network are the significant increase in observational sensitivity, and the availability of position information for the source of the gravitational radiation.
As discussed in \cite{Sylvestre}, a realistic goal for Advanced detectors is to trigger the observation of the electromagnetic radiation associated with compact binary mergers.

The simplest approach to analyzing data from a network of GW detectors is to process the data from each interferometer individually, and to combine the results incoherently.
Event lists are generated for every interferometer, and coincidence gates of various levels of complexity (which may include information on time, frequency, amplitude, etc.) are used to combine the lists into a single set of GW candidates. 
The source position can be estimated by triangulation when good information about the arrival time of the bursts in all interferometers is available.
This incoherent approach has been used for all the analyzes performed on real data to date \cite{thesis,burstanalyses}.

Recent theoretical work has demonstrated the existence of an optimal statistic for a {\it coherent} analysis when matched filtering is possible \cite{Finn,Pai}, i.e., when the phase evolution of the source is precisely known.
In essence, the coherent analysis combines the data {\em before} the search for the presence of a gravitational wave is performed.
This paper describes a parallel approach for the more complex case where no precise model of the gravitational wave signal is available.
Such signals are often detected in single interferometers using power detectors which rely on a quadratic form of the data as a detection statistics \cite{tfplane,tfplane1,epower,vicere,tfclusters,wnoise}.
I describe in section \ref{algorithm} an algorithm which optimally combines data from a network of interferometers into a single, {\it synthetic} data stream which can be processed by any of these single interferometer power detectors.
Section \ref{simulations} presents results for some special cases studied analytically or with numerical simulations.
A more complete discussion of the algorithm and of its performances is presented elsewhere \cite{paper1}, and is summarized in section \ref{discussion}.

\section{Algorithm} \label{algorithm}
Consider one of the many power detectors that were developed to detect GW bursts of arbitrary waveform in the data from individual interferometers.
It is assumed that the power detector produces a estimate $\hat{P}(\vc{y})$ of the power present in the signal for a given data vector $\vc{y}$, and that this estimated power can be expressed in terms of a scalar product,
\begin{equation}
\hat{P}(\vc{y}) = |\vc{y}|^2 = \vc{y} \cdot \vc{y}. \label{eq:power}
\end{equation}
One may try to recycle such a power detector for a network of $N$ independent interferometers by feeding it with a {\it synthetic} data vector $\vc{Y}$, formed by linearly combining the time-shifted data from the $N$ detectors:
\begin{equation}
\vc{Y} = \sum_{i=1}^N a_i T(\delta_i) \vc{y}_i. \label{eq:syntheticresponse}
\end{equation}
Here, $T$ is the time-shift operator, $\vec{a}$ is a set of scalar weights, and $\vec{\delta}$ is a set of time-shifts.

The cross-correlation function between the plus and the cross polarizations of the GW signal ($\vc{s}_+$ and $\vc{s}_\times$) is defined by
\begin{equation}
R_{+\times}(\Delta) = \vc{s}_+ \cdot T(\Delta)\vc{s}_\times.
\end{equation}
Under the condition that this function has an extremum at $\Delta=0$, i.e. that its derivative satisfies $R'_{+\times}(\Delta)|_{\Delta = 0} = 0$, it can be shown that the signal-to-noise ratio (SNR) for the network power, $\hat{P}(\vc{Y})$, is maximum when the time-shifts $\vec{\delta}$ are chosen to correspond to the true position of the source (as expected), and when the weights $\vec{a}$ are a solution of the following eigenvalue problem
\begin{eqnarray}
\lambda a_i \sigma_i^2 + \sum_{j=1}^N a_j [F^+_i F^+_j \vc{s}_+ \cdot \vc{s}_+ + (F^+_i F^\times_j + F^\times_i F^+_j) \vc{s}_+ \cdot \vc{s}_\times + F^\times_i F^\times_j \vc{s}_\times \cdot \vc{s}_\times] = 0, \label{eq:normalphaII}
\end{eqnarray}
where $i=1, ..., N$.
Here, $\lambda$ is an eigenvalue, $\sigma_i^2$ is the variance of $\vc{y}_i$, and $F^+_i$ and $F^\times_i$ are the interferometer beam-pattern functions for the two polarizations for the i$^{th}$ detector.

Consequently, when $R'_{+\times}(\Delta)|_{\Delta = 0} = 0$, the SNR can be maximized with respect to only four parameters: the position of the source $(\theta, \phi)$, the ratio of the RMS of the cross polarization to the RMS of the plus polarization ($\Lambda_{+/\times} = |\vc{s}_+| / |\vc{s}_\times|$) and the amount of ``overlap'' between the two polarizations ($\Lambda_{+\cdot\times} = \vc{s}_+ \cdot \vc{s}_\times / |\vc{s}_+||\vc{s}_\times|$).
It is very significant that the maximization can be performed over only four parameters; without this collapse of the parameter space from a space of extremely large dimensionality where the waveforms for the two polarizations are fully specified to a space with only four dimensions, no practical implementation of the method would be possible.
The coherent power filter ({\tt CPF}) algorithm described in this paper can search over all four parameters; it estimates the source parameters by returning the point in parameter space which provides the largest value of the SNR, as estimated from the data.
Explicitly, the {\tt CPF} algorithm is defined as follow:
\begin{enumerate}
\item Pick a set of trial source parameters $\theta$, $\phi$, $\Lambda_{+/\times}$, and $\Lambda_{+\cdot\times}$.
\item Compute $\vf{a}$ from Eq.(\ref{eq:normalphaII}), and set $\vf{\delta}=0$.
\item Form the synthetic response $\vc{Y}$ using Eq.(\ref{eq:syntheticresponse}).
\item Use a single interferometer power detector to calculate $\hat{P} = |\vc{Y}|^2$.
\item Retain the source parameters $\theta$, $\phi$, $\Lambda_{+/\times}$, and $\Lambda_{+\cdot\times}$ if $\hat{P}$ is the largest to date.
\item Go back to Step 1 unless all points in parameter space have already been visited.
\end{enumerate}

The fact that the problem can be reduced to a maximization over only four parameters can be understood by a simple geometrical picture.
Representing the time-series as vectors in a space of high dimensionality, the data observed by every interferometer are the sum of a random vector and of the linear combination of two signal vectors (for the plus and cross polarizations), appropriately rotated to account for the time of arrival of the signal at te detector.
The weights in this linear combination are just the beam-pattern functions for each interferometer.
If one picks a set of time-shifts that cancels out perfectly the differences in time of arrival, all the signal components of the time-shifted observation vectors will lie in a single hyperplane spanned by the two signal vectors.
The orientation of that hyperplane is unknown, since the signal vectors are unknown.
However, the goal is to maximize the norm of the weighted sum of the signal components of the time-shifted observation vectors.
This maximization can be performed, because this norm depends only on the beam-pattern functions, the angle between the two signal vectors (i.e., $\Lambda_{+\cdot\times}$), and the ratio of their norms (i.e., $\Lambda_{+/\times}$), up to an unimportant overall scale factor.

\section{Examples} \label{simulations}
The examples presented in this section are all for the network formed by the LIGO interferometer near Hanford, Washington, the LIGO interferometer near Livingston, Louisiana, and the Virgo interferometer (the HLV network).
Results for the study of a long, nearly monochromatic signal similar to a chirp from a binary inspiral are first discussed in relation to possible systematic errors on the position estimations performed by {\tt CPF}.
Results from a number of numerical simulations with a short, broader-band signal are then presented, in order to get insights into the statistics of the {\tt CPF} algorithm as a position estimator.

Obviously, the condition $R'_{+\times}(\Delta)|_{\Delta = 0} = 0$, which is assumed by the {\tt CPF} algorithm, cannot be expected to hold true in general, for an arbitrary GW signal.
I consider for instance a signal with characteristics similar to those of the chirp from a binary inspiral.
It is a long, monochromatic signal with a 40 Hz frequency and a $\pi/2$ phase offset between the plus and the cross polarization waveforms, so that $R'_{+\times}(\Delta)|_{\Delta = 0} \neq 0$.
A linearization of the normal equations for the SNR maximization problem can be used to show that the systematic error in position introduced by using the {\tt CPF} algorithm for this signal is rather large, i.e. is above $10^{-2}$ radian for 99\% of the sky.
On the other hand, the size of the second order systematic error is negligible ($\alt 10^{-2}$ radian) for approximately 25\% of the sky. 
Since these errors are deterministic, a knowledge of the value of $R'_{+\times}(\Delta)|_{\Delta = 0}$ is sufficient for implementing a correction procedure that removes the first order errors.
This correction is feasible for $\sim 25\%$ of the sky, thus limiting the sky coverage of the {\tt CPF} algorithm for source position estimation to one quarter of the sky, at any given time.
For signals less sensitive to systematic errors than the one described in the example above, the noise in the detectors will obviously dominate the errors in the position estimates.
Such errors are best studied using numerical simulations.

In all numerical simulations, the noise in the three interferometers was taken to be white noise of unit variance, and the data were processed in 10 s long segments sampled at 16 384 Hz.  
The plus and cross polarization waveforms were independent realization of a bandlimited white noise process, of duration 1/16 s, and with negligible power below 125 Hz and above 150 Hz.
The plus polarization waveform was produced from a broadband noise of variance 2, and the cross polarization from a broadband noise of variance 1/2, so that on average $\Lambda_{+/\times} = 2$.
Because of the independence of the two polarizations, on average $\Lambda_{+\cdot\times}$ was equal to zero.
In all cases, the signal was rather strong; the signal was scaled so that its optimal filtering amplitude network signal-to-noise ratio\footnote{i.e., the signal-to-noise ratio that could have been obtained has the arrival time, duration, and frequency band of the signal all been known beforehand} was 13.4.
All signals were injected from the northern hemisphere normal to the plane containing the three interferometers of the network.

\begin{figure}
\begin{center}
\ieps{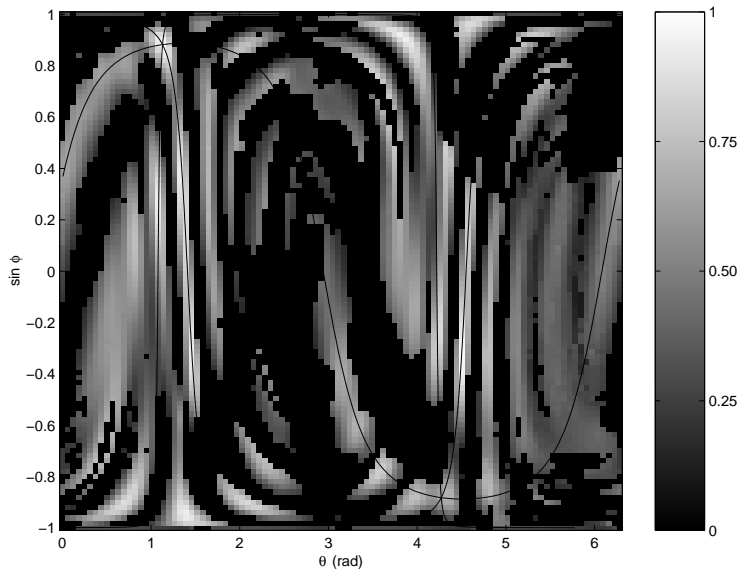}
\end{center}
\caption{An example of the output of the first pass of the {\tt CPF} algorithm. The position plane with the value of $\Lambda_{+\cdot\times}$ giving the maximal network power is plotted. The gray scale represents the network power, normalized to its maximum value. The continuous curves show the loci of equal time delays for the three independent interferometer pairs.}
\label{fig:posEx}
\end{figure}

\begin{figure}
\begin{center}
\ieps{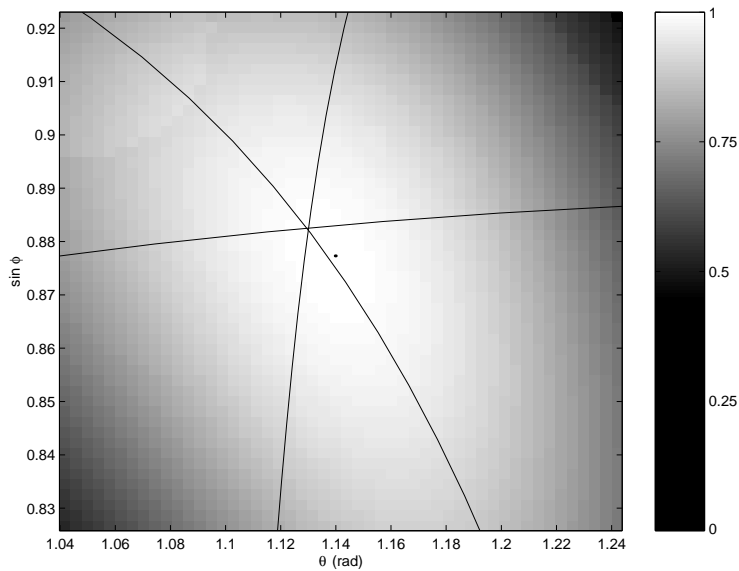}
\end{center}
\caption{Same as Figure \ref{fig:posEx}, for the second pass of the {\tt CPF} algorithm. The black dot shows the position with maximum network power.}
\label{fig:posExZ}
\end{figure}

All simulations were performed using the LIGO Data Analysis System (LDAS) implementation of the {\tt CPF} algorithm, which relies on the use of the {\tt TFCLUSTERS} algorithm as the single interferometer power detector which measures the network power.
In order to keep the size of the simulations at a manageable level, it was assumed that the value of $\Lambda_{+/\times}$ was known exactly, and this parameter was not included in the maximization of the network power.
For the other three parameters, the network power was first computed over 10 different values of $\Lambda_{+\cdot\times}$ uniformly distributed over the interval $[-1,1]$, 100 different values of the right ascension uniformly distributed over $[0, 2\pi[$, and 100 different values of the sine of the declination uniformly distributed in $[-1,1]$.
Consequently, one parameter scan of the {\tt CPF} algorithm involved $10^5$ runs of the {\tt TFCLUSTERS} algorithm, and took $\sim 225$ s on a cluster of 32 Pentium IV computers clocked at 2 GHz and equipped with 512 Mb of RAM.
In order to obtain a position estimate which was not limited by the resolution of the parameter space, a second run of the {\tt CPF} algorithm was performed.
For that second run, the value of $\Lambda_{+\cdot\times}$ was fixed to the value obtained in the first run, and both the right ascension and the sine of the declination were scanned with 50 points in each direction, in a square window of size 0.2 rad in both directions, centered on the position estimated in the first run.
Figures \ref{fig:posEx} and \ref{fig:posExZ} show an example of the two runs of {\tt CPF} for a single simulated data segment.

\begin{figure}
\begin{center}
\ieps{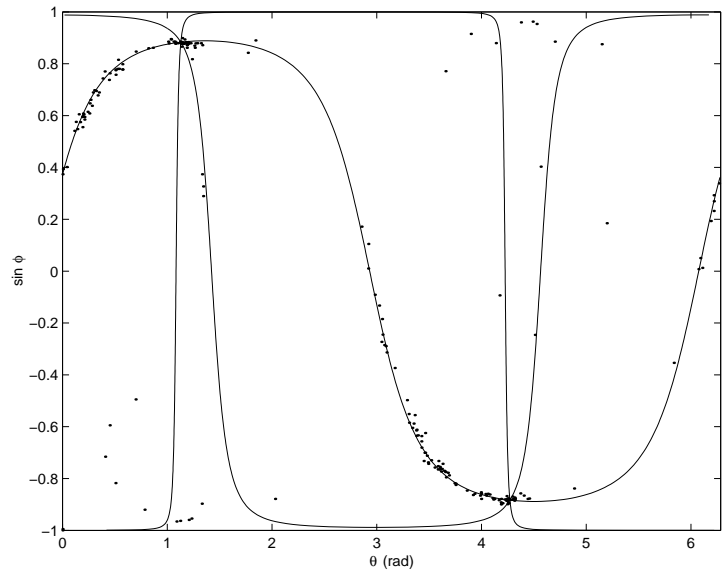}
\end{center}
\caption{Scatter plot of the estimated source position, for 240 realizations of the simulation.}
\label{fig:posU210}
\end{figure}

Running repeatedly the simulations gave a number of position estimates, which are plotted in figure \ref{fig:posU210}.
A significant number of points cluster around the position where the signal was injected, or around its mirror image with respect to the HLV plane, a natural consequence of the ambivalence present in any position estimation with only three interferometers.
Approximately 25\% of the trials had errors smaller than one degree, where the error is the arclength along the great circle connecting the estimate and the signal position or its mirror image, whichever is smaller.
Most of the other trials gave positions that were significantly off the source position or its mirror image, but which lied on the equal-delay curve for the Hanford-Livingston interferometer pair.
Physically, this results from the very good alignment of these two detectors, and the relatively poor alignment of the Virgo interferometer with them.
In particular, it can be seen that in addition to hugging the Hanford-Livingston equal delay curve, the position estimates gather into clusters along that line.
These clusters correspond to positions where the signals in the interferometers at Hanford and Livingston are in phase, and are at a phase offset with respect to the signal at Virgo which is an integer multiple of the characteristic wavelength of the signal.

\section{Discussion} \label{discussion}
There are significant difficulties inherent to the analysis and detection of GW bursts in the data from a realistic network of interferometers, such as the HLV network.
The complications come essentially from the entanglement of the two unknown polarizations of the GW signal, and are exacerbated when the two polarization waveforms do not show significant differences in their amplitudes or structures, or when the interferometers are poorly aligned.
A simple and optimally efficient way of mitigating these problems was described in this paper. 
It can be used to efficiently recycle burst detectors that have been shown to be efficient at picking signals in a given data stream, and that are robust and well characterized when operated on non Gaussian or colored noise.

Short bursts ($\sim 10$ cycles) have been studied numerically. 
For reasonable burst amplitudes, it was shown that their source could be located with good accuracy ($\sim 1^\circ$) a significant fraction of the time.
Variation of the parameters of the simulations \cite{paper1} revealed that the  position estimation error was scaling weakly with the burst amplitude, but was a stronger function of the degree of polarization of the wave, which was approximately controlled by the parameter $\Lambda_{+/\times}$.

The misalignment of the Virgo interferometer with respect to the two LIGO detectors in the HLV network was also a significant source of error in the estimation of the position.
In the simulations, the {\tt CPF} algorithm correctly placed the source at a position equidistant from the Hanford and from the Livingston LIGO detectors 90\% of the time.
It was, however, often unable to correctly use the information provided by the Virgo instrument.
The {\tt CPF} algorithm can be applied to networks of arbitrary complexity; it is likely that including other interferometers in the network will reduce the amount of points on the sky where no instrument can effectively complement the measurements from the two LIGO sites.

Although it has not been discussed in this paper, it should be noted that the {\tt CPF} algorithm can also be used to improve the detection efficiency of a network of interferometers.
It is shown in \cite{paper1} that {\tt CPF} achieves a higher probability of detection at the same false alarm rate than a incoherent approach based only on the comparison of event lists from the three interferometers of the HLV network.

\acknowledgments 
The author acknowledges the use of the computing ressources of the LIGO Scientific collaboration to carry out the simulations presented in this work.
This work was supported by the National Science Foundation under cooperative agreement PHY-9210038 and the award PHY-9801158.
This document has been assigned LIGO Laboratory document number LIGO-P030016-00-D.

\end{document}